\documentclass[prl,twocolumn,aps,showpacs]{revtex4}
\usepackage{graphicx}

\begin{document}

\title{Even Parity, Orbital Singlet and Spin Triplet Pairing for
Superconducting $La(O_{1-x}F_x)FeAs$}
\author{Xi Dai$^1$, Zhong Fang$^1$, Yi Zhou$^2$ and Fu-chun Zhang$^2$}
\affiliation{$^1$Beijing National Laboratory for Condensed Matter Physics, and Institute
of Physics, Chinese Academy of Sciences, Beijing 100190, China\\
$^2$Department of Physics, Center of Theoretical and Computational Physics,
the University of Hong Kong, Hong Kong, China}
\date{\today}

\begin{abstract}
In the present paper, we propose the parity even,orbital singlet and spin
triplet pairing state as the ground state of the newly discovered
super-conductor $LaO_{1-x}F_xFeAs$.The pairing mechanism involves both the
special shape of the electron fermi surface and the strong ferromagnetic
fluctuation induced by Hund's rule coupling.The special behavior of the
Bogoliubov quasi-particle spectrum may leads to "Fermi arc" like anisotropy
super-conducting gap, which can be detected by angle resolved photo
emission(ARPES).The impurity effects are also discussed.
\end{abstract}

\pacs{74.70.-b, 74.25.Jb, 74.25.Ha, 71.20.-b}
\maketitle

%
Recent discovery of superconductivity in layered $Fe$-based compounds has
attracted much attention. It has been reported that the transition
temperatures are $T_c=26$ K in $LaO_{1-x}F_xFeAs$\cite{LOFS} , $T_c=41$ K in
$CeO_{1-x}F_xFeAs$\cite{COFS} and $T_c=43$ K in $SmO_{1-x}F_xFeAs$\cite{SOFS}%
. This class of superconductors shows highly unusual properties, indicating
possible unconventional non-BCS superconductivity.\cite{NLWang,HHWen}

The electronic band structure calculations for $LaOFeAs$ suggest the
compound to be a semi-metal\cite{LOFP-cal,Singh,fang}. There is a perfect
nesting between the hole Fermi surface (FS) centered at $\Gamma$ point and
electron FS centered at $M$ point, which leads to a spin density wave state
at low temperature\cite{Mazin,SDW}. Superconductivity occurs when part of $%
Fe^{2+}$ ions are replaced by $Fe^+$, which removes the nesting. The layered
$Ni$ based compound is also superconducting (SC) although $T_c$ is much low%
\cite{LONS}. This implies the importance of ferromagnetic (FM) fluctuation
to the superconductivity. The similarity of $Fe$-based superconductors with $%
Sr_2RuO_4$ has suggested a possible spin triplet pairing. While the
conventional s-wave BCS state is robust against non-magnetic disorder due to
the Anderson theorem\cite{Anderson},the p-wave superconductivity of $%
Sr_2RuO_4$ is only observed in clean samples and is strongly suppressed by
the non-magnetic impurity\cite{BW,maeno}. The $Fe$-based superconductivity,
on the other hand, does not require a clean sample and appears to be robust
against disorder. Together with their high transitional temperatures, this
raises an important and interesting question on the symmetry of the newly
discovered $Fe$-based superconductivity.

In this Letter, motivated by the approximate two-fold degenerate electron FS
revealed in the electronic structure calculations, we propose a spin triplet
pairing with even parity for SC $LaO_{1-x}F_xFeAs$. The pairing is due to
the FM fluctuation between electrons in two different orbitals with almost
denegerate bands. Our theory explains the robustness of superconductivity to
the disorder in a spin triplet SC state, similiar to the disorder effect to
the spin singlet s-wave BCS superconductor. The splitting of the orbital
denegeracy strongly suppresses the superconductivity. The high pressure
reduces the splitting, and may further increase $T_c$ in $LaO_{1-x}F_xFeAs$.
The splitting of the degeneracy also leads to a pronounced $\vec k$%
-dependence in the isotropic s-wave SC state, which may be tested
in angle resolved photoemission spectra (ARPES).

We start from the special electronic structure of layered compound $LaOFeAs$%
. The $Fe$-ions forms a square lattice with two atoms in each unit cell. The
distance of neighboring $Fe$ atoms is rather short, so that the electron
direct hoppings between $Fe$ ions are important, similar to the elemental $%
Fe $, which is a FM metal. Due to the multiple $d$-orbitals, there are five
FS with three hole-like cylinders around the $\Gamma$ point and two
electron-like cylinders around the $M$ point of the Brillouine zone. Upon
doping of $F$-atoms, the three hole-like FS shrink rapidly, while the two
electron FS expand their areas. Therefore, it is reasonable to expect that
the two bands of electron-like states are responsible for the
superconductivity in $La_{1-x}F_xOFeAs$. Competing spin fluctuations exist
in this compound. One is the anti-ferromagnetic spin fluctuations due to the
nesting between the electron and the hole FS, which are connected by a
commensurate $q$-vector. The other is the FM spin fluctuation likely due to
Hund's coupling. The presence of the nesting between electron and hole FS
will induce a spin density wave instability. This is the $150 K$ anomaly
observed experimentally\cite{NLWang}. Doping $F$-ions destroys the spin
density wave state and opens the door for superconductivity.

The itinerant ferromagnetism is an interesting but difficult problem in the
condensed matter theory with a long history\cite{ferro1,ferro2,ferro3}. One
of the important issues is if the multi-band nature is necessary for the
itinerant ferromagnetism. Both the analytical and numerical studies indicate
that itinerant ferromagnetism is very difficult to obtain in the single band
system, unless the Fermi energy is close to a van hove singularity\cite%
{ferro1,ferro2}. Usually the multi-bands are necessary to stabilize the FM
phase, and the Hund's rule coupling plays a crucial role. In $LaOFeAs$, the
density of state is very low near the FS, as evidenced in both optical
conductivity measurement\cite{SDW} and first principle calculation by
several groups\cite{Singh,fang}. Therefore, the Hund's rule coupling is
likely to be the main reason for the FM fluctuation here. It is thus
reasonable to speculate that the pairing glue of the superconductivity in $%
LaO_{1-x}F_xFeAs$ be the inter-band FM fluctuation and the Cooper
pair is formed by the spin triplet pairs of the electrons on two
different bands.

In what follows we examine the SC properties of a model Hamiltonian
consisting of two approximately degenerate bands denoted by orbital $1$ and
orbital $2$ and a pairing field between the two orbitals with parallel
spins. The band structure of the model reproduces the two electron bands
obtained in the local density approximation for $LaO_{1-x}F_{x}FeAs$. The
pairing interaction is from the inter-band ferromagnetic fluctuation indued
by the Hund's coupling. We consider a tight binding model for electrons $C_{%
\vec{k}\sigma \alpha }$ in a square lattice given by

\begin{equation}
H=\sum_{k\sigma \alpha }\left( \varepsilon _{k,\alpha }-\mu \right)
C_{k\sigma \alpha }^{\dag }C_{k\sigma \alpha }-J_{k-k^{\prime
}}\sum_{kk^{\prime }m}\widehat{\Delta }_{km}^{\dag }\widehat{\Delta }%
_{k^{\prime }m}
\end{equation}%
where $\alpha =1,2$ are orbital indicies, and
\begin{eqnarray}
\varepsilon _{k,1} &=&t\gamma _{k}+t_{1}\gamma _{k}^{(1)}+t_{2}\gamma
_{k}^{(2)},  \nonumber \\
\varepsilon _{k,2} &=&t\gamma _{k}+t_{2}\gamma _{k}^{(2)}+t_{1}\gamma
_{k}^{(1)},
\end{eqnarray}

with $\gamma _{k}=\cos {k_{x}}+\cos {k_{y},}\gamma _{k}^{(1)}=\cos
(k_{x}+k_{y})$ and $\gamma _{k}^{(2)}=\cos (k_{x}-k_{y})$.

In the calculations below, we choose $t=0.3eV$, $t_{1}/t=0.267$, which are
obtained by approximately fitting the shape of the two electron FS and the
overall band width with the first principle calculations for $LaOFeAs$\cite%
{fang}. We consider $t_{2}$ as a tuning parameter to study the effect of the
FS anisotropy, with $t_{2}/t_{1}=1$ corresponding to the isotropic case, and
$t_{2}/t_{1}=0.6$ for undoped $LaOFeAs$, $t_{2}/t_{1}=0.8$ for $%
LaO_{0.9}F_{0.1}FeAs$  under the normal pressure. $\mu $ is the chmecial
potential.

The second term in $H$ describes an inter-band pairing interaction with $J_{%
\vec{k}}$ the pairing strength, and $m=1,0,-1$ the three components in the
spin triplet state.
\begin{eqnarray}
\widehat{\Delta }_{k,1}^{\dag } &=&C_{k\uparrow ,1}^{\dag }C_{-k\uparrow
,2}^{\dag },  \nonumber \\
\widehat{\Delta }_{k,-1}^{\dag } &=&C_{k\downarrow ,1}^{\dag
}C_{-k\downarrow ,2}^{\dag },  \nonumber \\
\widehat{\Delta }_{k,0}^{\dag } &=&1/\sqrt{2}\left( C_{k\downarrow ,1}^{\dag
}C_{-k\uparrow ,2}^{\dag }+C_{k\uparrow ,1}^{\dag }C_{-k\downarrow ,2}^{\dag
}\right) .
\end{eqnarray}%
We note that the spin triplet Cooper pairs described by $\Delta $ above are
singlets in orbital sector. Therefore,in order to obey the Fermi statistics,
the spatial part of the wave function must be of even parity, such as $s$%
-wave, or extended $s$-wave, or $d$-wave.

We now turn to the discussion of the inter-band pairing strength $J_{\vec{k}%
} $. In general $J_{\vec{k}}$ can be expanded by crystal harmonics as
\begin{equation}
J_{k}=J_{0}+J_{1}(\cos {k_{x}}+\cos {k_{y}})+...;  \label{hund}
\end{equation}%
where $J_{0}$ may be viewed as the on-site Hund's coupling between two
Wannier orbitals centered on the same site and $J_{1}$ is the magnetic
coupling between the neighboring sites, which may be induced by the Coulomb
exchange interaction between the Wannier orbitals in the itinerant electron
systems. If $J_{0}$ is strong enough to overcome the on-site direct Coulomb
interaction $U$ between the two orbitals, an s-wave spin-triplet pairing
state is favored. If $J_{1}$ is large or if $U$ is large, extended s-wave or
d-wave pairing states could be stablized to avoid the cost in $U$. Note that
the possibility of the spin triplet pairing state induced by FM fluctuation
was previously proposed to explain the superconductivity in $Sr_{2}RuO_{4}$%
\cite{triplet} and the inter-band pairing was discussed by a number of
authors previously\cite{YP,Tosatti}.

\begin{figure}[t]
\includegraphics[width=4cm,angle=0,clip]{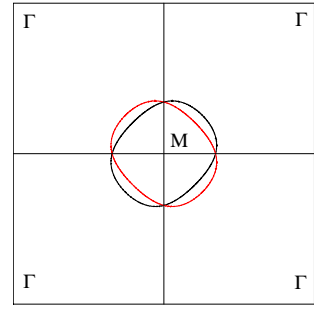}
\caption{The Fermi surface (FS) derived from the two-band tight
binding model $H$ of Eq. (1) in the absence of the pairing, which
reproduces well the electron FS obtained by first principle
calculation\protect\cite{fang} for undoped $LaOFeAs$. }
\label{fig1}
\end{figure}

Before we discuss the properties of the super-conducting phase, we first
examine the special shape of the Fermi surfaces obtained by the above
Hamiltonian with the electron filling $\delta =0.08$ \ for each band, which
are shown in Figure \ref{fig1}. The two Fermi surfaces match each other
perfectly by rotating with 90 degrees and the degeneracy along the M-X line
is guaranteed by the four-fold rotational symmetry. Because of the metallic
nature of the system, the crystal anisotropy is not so strong, and the two
Fermi surfaces overlap to each other quite well, which gives the system
relatively large phase space for the inter-band pairing.

The above Hamiltonian can be solved by mean field decoupling assuming that
only $\sum_{k^{\prime }}J_{k-k^{\prime }}\left\langle \widehat{\Delta }%
_{k^{\prime }0}\right\rangle =\Delta _{k,0}\neq 0$ and $\Delta _{k,1}=\Delta
_{k,-1}=0$, which is the natural choice for the phase with time reversal
symmetry. Therefore, the mean field Hamiltonian reads,

\bigskip
\[
H_{mf}=\Psi _{k}^{\dag }\left(
\begin{array}{cc}
\widehat{h}_{k} & 0 \\
0 & \widehat{h}_{k}%
\end{array}%
\right)
\]

with 2$\times $2 matrix $\widehat{h}_{k}=-\delta
_{k}\widehat{1}+\left( \varepsilon _{k,1}-\mu +\delta _{k}\right)
\widehat{\sigma }_{z}+\Delta _{k,0}\widehat{\sigma }_{x}$, $\delta
_{k}=1/2\left( \varepsilon _{-k,2}-\varepsilon _{k,1}\right) $ and
$\Psi _{k}^{\dag }=\left( C_{k\uparrow ,1}^{\dag },C_{-k\downarrow
,2},C_{k\downarrow ,1}^{\dag },C_{k\uparrow ,2}\right) $ to be the
Nambu representation. The Bogoliubov quasi-particle spectrum can
be obtained by solve the above Hamiltonian, which can be written
as $E_{k\sigma \pm }=-\delta _{k}\pm \sqrt{\left( \varepsilon
_{k,1}-\mu +\delta _{k}\right) ^{2}+\Delta _{k,0}^{2}}$. The above
Bogoliubov quasi-particle spectrum  gives the minimum gaps sizes
detected by angle resolved photo emission to be $E_{\min
}^{gap}=\max (0,|\Delta _{0}\left( \theta \right) |-|\delta
_{k_{F}}\left( \theta \right) |)$, where $\theta $ denotes the
angle around the Fermi surface. \ The band splitting $\delta
_{k_{F}}$ has strong angle dependence, which vanishes at the four
crossing points with $\theta =\frac{n\pi }{2}$and reaches the
maximum at $\theta =\frac{(2n+1)\pi }{4}$. Thus even the order parameter $%
\Delta _{k0}$ itself is isotropic, i.e. the s-wave or extended s-wave case,
the super-conducting gap can have strong angle dependence if $\delta _{k_{F}}
$ is compatible to $\Delta _{k,0}$. Further, if the absolute value of $%
\delta _{k_{F}}$ is bigger than the amplitude of the order parameter $\Delta
_{k,0}$ at some specific angle, there will be no gap at the FS along that
direction. Therefore a "Fermi arc" may appear. If the spacial pairing
symmetry is s-wave or extended s-wave, the "Fermi arc" only appears for very
small order parameter $\Delta _{k,0}$. While for d-wave case, since $\delta
_{k_{F}}\left( \theta \right) $ takes the maximum value along the d-wave
nodal direction where the order parameter vanishes, the "Fermi arc" will be
always there.

\begin{figure}[t]
\includegraphics[width=4cm,angle=0,clip]{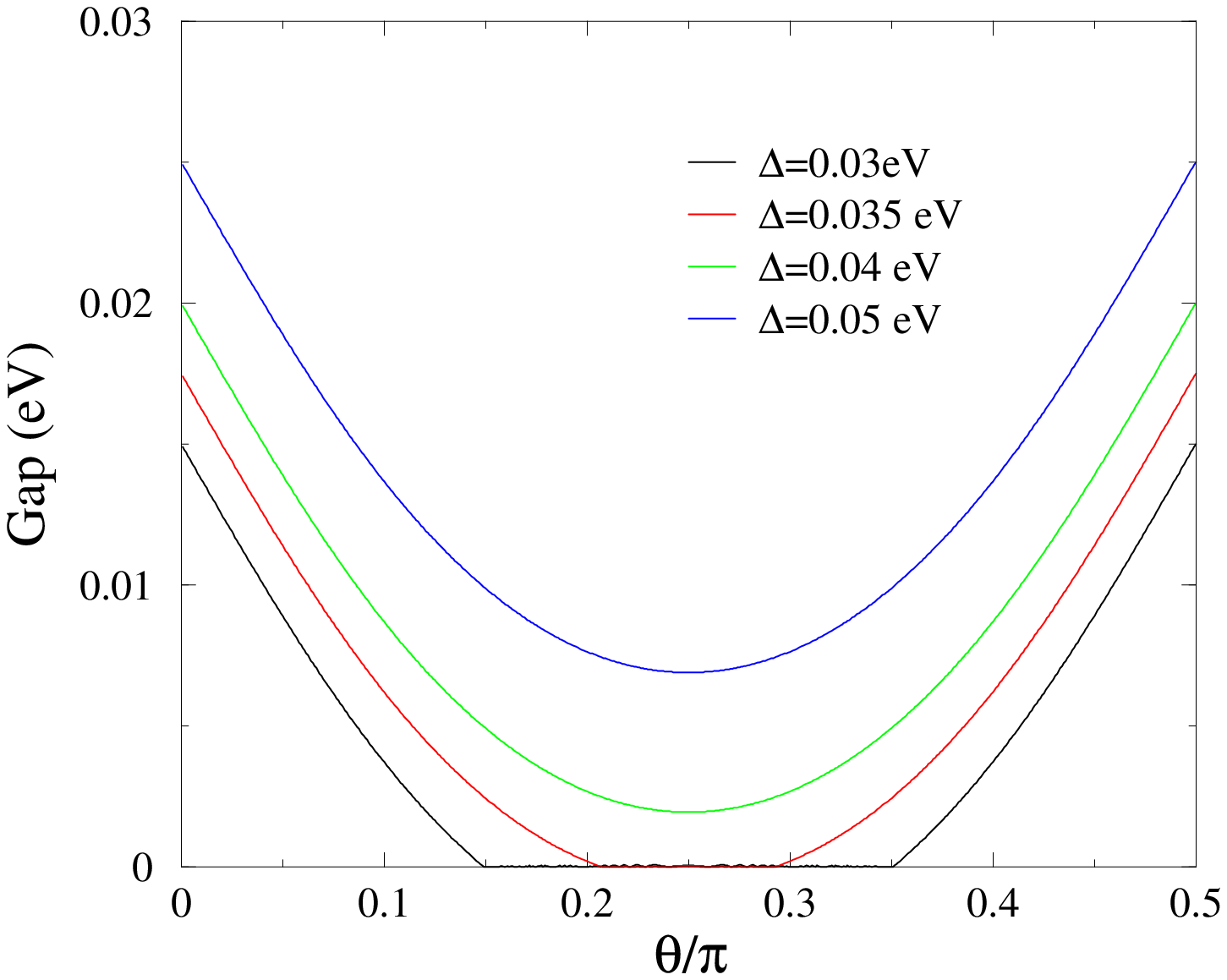} %
\includegraphics[width=4cm,angle=0,clip]{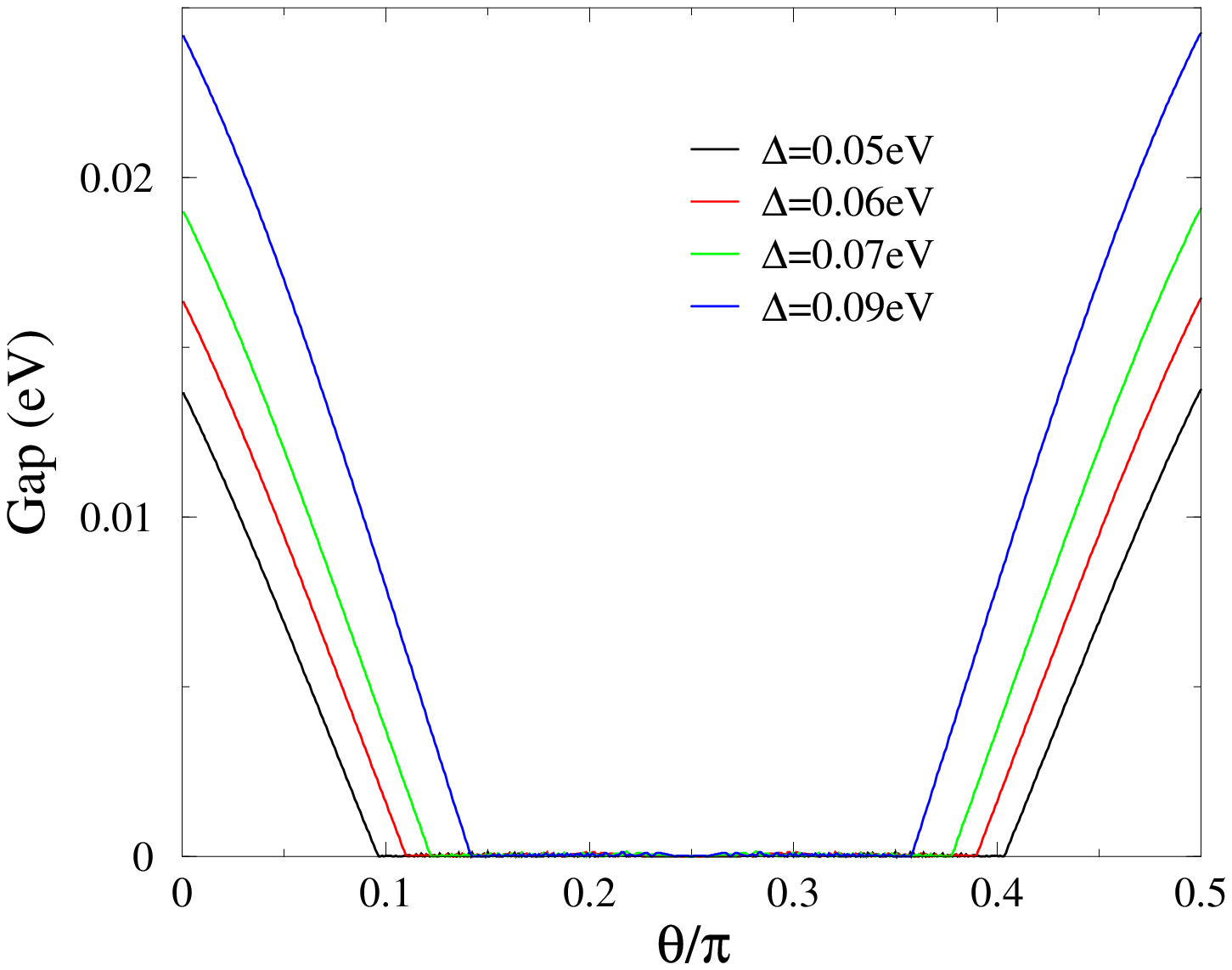}
\caption{The angle dependence of the minimum gap value at the FS
with four different value of the super-conducting order
parameter.} \label{fig2}
\end{figure}

In Fig. \ref{fig2}, we plot the angle dependence of the gap function on the
Fermi surface with four different values of the order parameter for both
s-wave (a) and d-wave case (b). For the s-wave case, the "Fermi arc" appears
only when the order parameter is small. While for d-wave case, it always
exists. The strange behavior of the Bogoliubov quasi-particles indicate that
it is possible to have low lying excitations in this orbital singlet, spin
triplet state even with s-wave or extended s-wave pairing symmetry.

\begin{figure}[t]
\includegraphics[width=5cm,angle=0,clip]{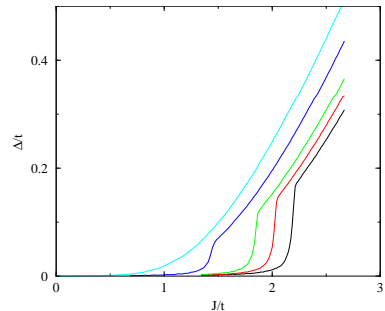}
\caption{The super-conducting order parameter as the function of effective
ferromagnetic exchange coupling $J_0$. From left to right, the tight binding
parameter $t_2$ equal to 0.08,0.06,0.04,0.03,0.02 eV respectively.}
\label{fig3}
\end{figure}

Note that there are four points on the two Fermi surfaces to match
exactly. The Cooper instability does not occur with infinitesimal
coupling $J$. Instead there is a quantum phase transition with
critical value of $J_{c}$, above which the inter-band spin triplet
pairing state has lower energy. First we assume the on-site
inter-orbital repulsion is not strong enough to suppress the
on-site triplet pairing. In that case, we only consider the
on-site term of the effective paring strength $J_{0}$ and applied
a mean field theory in s-wave channel to solve the Hamiltonian and
calculate the super-conducting order parameter $\Delta _{0}$ as a
function of $J$ for various ratios of $t_{2}/t_{1}$,
characteralizing the crystal anisotropy. The results are plotted
in Fig.\ref{fig3}. As we can see, the critical $J_{c}$ depends
strongly on the crystal anisotropy. For
the case in $LaOFeAs$, where $t_{2}/t_{1}$ is around $0.8$, the critical $%
J_{c}$ is found to be around $0.4eV$, which is quite feasible for iron
compounds. Our mean field theory suggests that the high sensitivity of the
super-conducting gap hence the transition temperature $T_{c}$ to the
anisotropy or the deviation of the approximately degenerate bands. $T_{c}$
may be raised dramatically if the anisotropy is reduced. We speculate the
high pressure measurement may reduce the anisotropy and hence increase $%
T_{c} $.

\bigskip

\bigskip When the on-site repulsion is strong, the on-site inter-orbital
triplet pairing will be surpressed. In that case, the nearest neighbor
Hund's coupling $J_{1}$ will be important and the spacial paring symmetry
may be extended s-wave or d-wave. For the $LaO_{1-x}F_{x}FeAs$ compounds,
from the LDA calculation the effective filling factor for the two electron
pocket is around $10\%$, which strongly favors the extended s-wave against
the d-wave pairing. While if the effective filling factor is increased by
either further doping the system or correlation effect, the d-wave pairing
state may also be stabilized.

Below we examine the impurity effect to the proposed pairing state. As it is
well known, in the absence of the orbital degrees of freedom, we have even
parity with spin singlet or odd parity with spin triplet. the spin singlet
s-wave superconductivity is unaffected by nonmagnetic impurities due to
Anderson's theorem\cite{Anderson}, but is strongly affected by magnetic
impurities\cite{BW}. On the other hand, a p-wave superconductor with spin
triplet is very sensitive to both non-magnetic and magnetic impurities\cite%
{BW}. This explains why spin triplet p-wave superconducting state $%
Sr_{2}RuO_{4}$ requires clean sample. For the orbitally paired state, the
impurity effect to the spin triplet state is very different.

We consider the proposed even parity, orbital singlet and spin triplet
state. We shall focus on the s-wave pairing. The case for extended s-wave
case will be similar. We follow Balian and Werthamer to apply a perturbation
theory to calculate the change of the free energy due to the impurity for
the proposed state. In the weak coupling limit, the change of free energy is
$\delta(F_s-F_n)\propto (1-\gamma)$ due to impurity scattering, where $%
\gamma $ is a coherence factor determined by both the impurity scattering
and the super-conducting state. $\gamma=1$ coresponds to vansihing effect,
while $\gamma=-1$ coresponds to the strongest suppression. In the
conventional pairing state, an s-wave state scattered by nonmagnetic
impurities leads to $\gamma=1$, hence the change of free energy is zero at
leading order, while such an s-wave state scattered by magnetic impurities
will result in $\gamma=-1$, indicating a very strong suppression. A p-wave
state scattered by either nonmagnetic or magnetic impurities will lead to $%
\gamma=0$ by average over k-space, indicating strong suppression. We have
found that for the proposed orbital singlet state, the s-wave with spin
triplet pairing state has $\gamma=1$ for nonmagnetic impurities, and $%
\gamma=1/3$ for the magnetic impurities. Therefore the state is robust
against non-magnetic impurity and is relatively weakly suppressed by
magnetic impurity.

 We may speculate the effect of "orbital impurity" which
flip the orbitals as the magnetic impurity flips the spin. Such orbital
impurities would strongly suppress the proposed superconductivity in a way
similar to the magnetic impurity to suppress conventional s-wave state.
While since the orbital degeneracy along the M-X line (which is crucial for
inter-band pairing) is protected by four-fold rotational symmetry within the
Fe-As plane, impurities which do not break the local four-fold symmetry
(such as the off plane impurities) will only generate very weak "orbital
flip" scattering terms. Therefore for this system, the off-plane impurities
act like the non-magnetic impurities in the traditional spin singlet
super-conductor, which has very little effect for s-wave or extended s-wave.
While the in-plane impurities, which will induce the local lattice
distortion and generate the inter-band scattering, act like the magnetic
impurities for the traditional spin singlet super-conductor, which will kill
the superconductivity very efficiently.

In summary, in the present letter we have proposed the parity even, orbital
singlet but spin triplet pairing state for the newly discovered
super-conductor $LaO_{1-x}F_xFeAs$. The pairing glue of the SC phase is the
strong ferromagnetic fluctuation induced by the Hund's rule coupling in the
iron compound. The pairing state is insensitive to the non-magnetic disorder
in contrary to the p-wave spin-triplet state. The Bogoliubov quasi-particle
spectrum has quite different behavior with the conventional s-wave spin
singlet super-conducting phase, which leads to the possible anisotropy in
the gap function and can be detected by angle resolved photoemission
spectral.

We acknowledge valuable discussions with Y. P. Wang, N. L. Wang, and the
supports from NSF of China and that from the 973 program of China, and RGC
grant of HKSAR.

\end{document}